\documentclass[10pt,conference,letterpaper]{IEEEtran}
\IEEEoverridecommandlockouts
\usepackage{cite}
\usepackage{amsmath,amssymb,amsfonts}
\usepackage{algorithmic}
\usepackage{graphicx}
\usepackage{eso-pic}
\usepackage{textcomp}
\usepackage{xcolor}
\usepackage{booktabs}
\usepackage{array}
\usepackage{longtable}
\usepackage{multirow}
\usepackage{tabularx}
\usepackage{gensymb}
\usepackage[hyphens]{url}
\usepackage{xurl}
\usepackage{rotating}
\usepackage[euler]{textgreek}
\usepackage{placeins}
\usepackage{float}
\usepackage{makecell}
\usepackage[colorlinks=true, allcolors=blue]{hyperref}

\def\BibTeX{{\rm B\kern-.05em{\sc i\kern-.025em b}\kern-.08em
    T\kern-.1667em\lower.7ex\hbox{E}\kern-.125emX}}

\newcommand{\takeaway}[1]{%
    \par\smallskip\smallskip
    \noindent\begingroup
    \setlength{\fboxsep}{4pt}%
    \setlength{\fboxrule}{0.6pt}%
    \fcolorbox{blue!40!black}{blue!4}{%
        \begin{minipage}{0.97\linewidth}
            \textcolor{blue!50!black}{\textbf{Takeaway.}} #1
        \end{minipage}%
    }%
    \endgroup
    \par\smallskip
}

\AddToShipoutPictureFG*{%
    \AtPageLowerLeft{%
        \raisebox{50pt}{%
            \makebox[\paperwidth][c]{%
                \resizebox{0.9\paperwidth}{!}{Preprint. Published at the 34th International Symposium on the Modeling, Analysis and Simulation of Computer and Telecommunication Systems (MASCOTS 2026).}%
            }%
        }%
    }%
}

\begin{document}

\title{Evaluating MFU as a Proxy for GPU Power for Energy-Aware Simulation of LLM Training}
% Evaluating MFU as a Proxy for GPU Power for Energy-Aware Simulation of LLM Training

\author{\IEEEauthorblockN{Niklas Enskat}
\IEEEauthorblockA{\textit{Technische Universität Berlin} \\
enskat@campus.tu-berlin.de}
\and
\IEEEauthorblockN{Philipp Wiesner}
\IEEEauthorblockA{\textit{Technische Universität Berlin} \\
wiesner@tu-berlin.de}
}

\maketitle

\begin{abstract}
High-fidelity performance simulators are essential for designing and configuring efficient AI systems, yet today's tools lack the ability to predict power consumption.
Established GPU power models rely on hardware utilization counters, which do not exist until the workload has actually run.
This work evaluates whether Model FLOPs Utilization (MFU)---an analytical, software-defined metric relating achieved throughput to peak hardware capability---can serve as a portable, software-defined predictor of GPU power for LLMs.
We benchmark almost 3000 single-device training runs across six GPUs, covering different model families, numerical precisions, batch sizes, and context-window lengths.
We find that a linear MFU-based power model fits every tested GPU as long as the workload is compute-bound, as in production LLM training.
Fitting per-(GPU, dtype, batch size) instead of per-GPU drops the within-cell mean error from around 10\,\% to around 1\,\%, matching the cross-repeat measurement-noise floor.
\end{abstract}


\section{Introduction}
The rapid progress of artificial intelligence (AI) has led to increasingly larger and more energy-demanding models~\cite{yang-2024B}.
Training and deploying such models relies heavily on GPU-accelerated infrastructures, where GPUs have become the backbone of modern AI systems~\cite{jahanshahi-2020}.
As AI workloads contribute a growing share of global data-center electricity use~\cite{international-energy-agency-2025}, there is an intensifying need for energy-aware system design and predictive performance modeling~\cite{katsenou-2024, google-2024}.
High-fidelity simulators and analytical models are the primary tools used by researchers and system architects to explore "what-if" scenarios for future AI clusters, such as optimizing scheduling policies or evaluating different hardware parallelization strategies~\cite{agrawal2024vidur, simumax, zhebrak2026llmclustersim}.

However, predicting the power consumption of simulated AI workloads remains a fundamental challenge.
Existing power models typically rely on hardware-level telemetry, such as GPU Utilization, which represents an aggregate measure of device activity reported by the hardware during execution~\cite{antarctica-2025, bridges-2016}.
While useful for operational monitoring, these metrics present a telemetry gap for simulation and performance modeling.
First, hardware-defined counters are only available \emph{after} code execution on physical hardware, making them inaccessible during the design or simulation phase.
Second, these metrics are vendor-specific (e.g., NVIDIA's NVML vs. AMD's ROCm), limiting the portability of derived power models across different providers or architectural generations.

A recently introduced yet already widely adopted performance metric for generative AI systems is Model FLOPs Utilization (MFU)~\cite{chowdhery-2023}.
Unlike hardware counters, MFU is an analytical metric that relates observed throughput to the theoretical peak compute capacity required per token.
Because MFU is derived from the model structure and high-level execution parameters, it can be calculated by simulators (e.g., Microsoft's Vidur~\cite{agrawal-2025}) without access to physical hardware.
While MFU has been proposed as a candidate input for power modeling in these settings~\cite{ozcan-2025}, its predictive value and robustness across diverse hardware have not been systematically evaluated.

This paper investigates whether MFU can serve as a reliable, portable, and software-defined proxy for GPU power consumption.
For LLM training, which is predominantly compute-bound, MFU captures the compute saturation that drives power consumption.
We validate MFU against ground-truth hardware measurements, providing a portable methodology for energy-aware simulation without vendor-specific telemetry.

To do so, we build a cross-architectural benchmarking framework and execute a controlled single-GPU training sweep on \emph{six} accelerators spanning two vendors and four hardware tiers (NVIDIA A100, L40, L4, Quadro RTX~5000, RTX~4070~Ti, and AMD MI210).
The sweep crosses three model families (Qwen2.5, GPT-2, DialoGPT), three numerical precisions (fp32, fp16, bf16), seven batch sizes (1--128), and two context-window lengths (512, 2048~tokens), yielding 126 configurations per device, each repeated four times with telemetry sampled until statistical convergence.
For every configuration, we record MFU, vendor-reported GPU~Utilization, and GPU power, and use this dataset to evaluate MFU and GPU~Utilization as predictors of power across hardware platforms and workload regimes.
We use internal GPU telemetry (NVML on NVIDIA, ROCm on AMD) as the reference power signal throughout.
Appendix~\ref{sec:power-validation} documents agreement with external wall-power measurements for two devices.

Our key findings are:
\begin{itemize}
    \item A linear MFU-power model fits every GPU we measure, including AMD MI210 where vendor-reported GPU~Utilization is a binary activity flag and admits no linear fit at all.
    \item The linear relationship holds across vendors and precisions, but each device needs its own fit because MFU is normalized by peak FLOPS. Supporting a new accelerator therefore takes only a one-shot per-GPU calibration sweep.
    \item Conditioning a separate linear fit per (GPU, dtype, batch size) collapses MAPE from $\sim$10\,\% per-GPU to $\sim$1\,\% per cell, which is statistically indistinguishable from the cross-repeat measurement-noise floor. The practical recipe for an MFU-based power proxy is therefore to condition on these three axes and on nothing else.
    \item At batch~1, MFU collapses below 1.3\,\% while power stays at $\sim$60\,\% of the batch-128 level. Further-out memory-bound regimes need a complementary memory-bandwidth signal.
\end{itemize}

Our benchmarking pipeline and data are publicly available: \url{https://github.com/dos-group/gpu_power_benchmark}.

\section{Performance and Power Metrics for GPUs}\label{Background}
Modern LLM training workloads impose sustained high computational demand and correspondingly high GPU power draw, so understanding how workload performance metrics relate to energy behavior is central to both efficient operation and predictive modeling.
Floating-point operations per second (FLOPS) are the conventional starting point for describing GPU compute capability, with vendors reporting theoretical peaks under idealized conditions.
Real training workloads rarely approach these peaks because of memory effects, scheduling overhead, communication latency, and kernel structure, so the relevant practical question is which fraction of peak compute a workload actually realizes and how that fraction relates to power.
The remainder of this section reviews two utilization metrics commonly used to answer this question (Section~\ref{sec:bg-gpu-util},~\ref{sec:bg-mfu}) and then outlines how GPU power is measured and estimated in practice (Section~\ref{sec:bg-power}).

\subsection{GPU Utilization}\label{sec:bg-gpu-util}
GPU Utilization is commonly reported via tools such as \texttt{nvidia-smi} or \texttt{rocm-smi} and is typically defined as the fraction of time during which the GPU is actively executing kernels within a sampling interval~\cite{cornelius-2025, nvidia-no-dateA}.
It is widely used for operational monitoring due to its simplicity and availability.
However, GPU~Utilization is a time-based activity metric rather than a measure of compute-efficiency.
High utilization may correspond to memory-bound execution, synchronization overhead, or stalled kernels, without implying productive throughput.
It is also a vendor-defined counter whose semantics differ across vendors and architectural generations, so equal values can correspond to substantially different throughput and power behavior.
These characteristics limit GPU~Utilization as a universal efficiency metric and raise the question about its suitability as a predictor of energy consumption.
It is nevertheless frequently used as an input to GPU power models~\cite{antarctica-2025, bridges-2016}.

\subsection{Model FLOPs Utilization (MFU)}\label{sec:bg-mfu}
MFU relates observed throughput (tokens per second) to the theoretical compute required per token, normalized by the peak compute capacity of the GPU for the specific numerical precision (dtype) used.
Unlike a hardware-defined activity counter, MFU is derived purely from the model architecture and measured throughput, making it a software-defined metric that can be computed inside a simulator without access to physical hardware.
Concretely, MFU~\cite{chowdhery-2023,ray-2025} is defined as
\begin{equation*}
    \text{MFU} = \frac{\text{Tokens/s} \cdot C_{\text{req}}}{\text{FLOPS}_{\text{peak}}},
\end{equation*}
where $C_{\text{req}}$ is the floating-point \emph{operation count} required for a forward and backward pass of a single token, and $\text{FLOPS}_{\text{peak}}$ is the theoretical peak throughput (operations per second) for the used precision and hardware generation.
For dense transformer models, $C_{\text{req}}$ is commonly approximated as
\begin{equation*}
    C_{\text{req}} = 6N + 12\,L_{\text{num}} H_{\text{num}} Q_{\text{dim}} T_{\text{seq}},
\end{equation*}
where $N$ is the number of model parameters, $L_{\text{num}}$ the number of layers, $H_{\text{num}}$ the number of attention heads, $Q_{\text{dim}}$ the head dimension, and $T_{\text{seq}}$ the sequence length~\cite{chowdhery-2023}.
In practice the exact operation count is more involved due to architectural and implementation details, so we use the integrated profiler CalFLOPS~\cite{calflops} to obtain $C_{\text{req}}$ directly.

Whether MFU can act as a reliable predictor of GPU power consumption depends on the execution regime.
In \textit{compute-bound} regions, performance is limited by arithmetic throughput, which MFU captures by construction.
In \textit{memory-bound} regions, the bottleneck shifts to data transfer between memory and processors, so GPU~Utilization may remain high while MFU drops.
A linear MFU--power relationship is therefore not expected to hold in this regime, and we restrict our study to the compute-bound case typical of production LLM training.
Because MFU reflects effective training progress rather than total executed operations, it is also less sensitive to implementation-specific recomputation effects than counter-based metrics.
This makes MFU an attractive abstraction for comparing training efficiency across devices and configurations.
Yet, as a power predictor, it has not been systematically validated, motivating the study in the remainder of this paper.

\subsection{GPU Power Measurement and Estimation}\label{sec:bg-power}
Modern data-center and workstation GPUs expose on-board power telemetry through vendor management interfaces (NVML on NVIDIA, ROCm SMI on AMD), reporting board-level power at sub-second granularity~\cite{weakley-2025}.
These signals are convenient and ubiquitous, and we use them as the ground-truth power signal throughout this study.
Appendix~\ref{sec:power-validation} validates them against external node-level wall-power measurements for two of our devices.
External hardware power meters can provide reference measurements at higher accuracy and at component or node level~\cite{yang-2024A, jay-2023}, but their deployment requires dedicated instrumentation and does not scale easily to large clusters.
Where direct measurement is impractical (most prominently in simulation and design-space exploration) analytical and statistical models estimate power from hardware activity counters and performance metrics~\cite{lang-2013}.
Such models depend on calibration data and typically rely on architecture-specific telemetry, which limits their portability across vendors and generations~\cite{yang-2024A} and prevents their use in simulators that have no hardware counters to read in the first place.
This portability gap is exactly what an MFU-based power model aims to close.

\section{Experimental Framework}\label{methodology}
To validate MFU as a simulation proxy, we develop a cross-architectural benchmarking framework that maps high-level model parameters to empirical power consumption.
We organize the framework into the workloads we run (Section~\ref{sec:workloads}), the protocol used to measure them (Section~\ref{sec:execution}), and the analysis pipeline that turns the raw logs into the regressions reported in Section~\ref{results} (Section~\ref{sec:analysis}).

\subsection{Workloads}\label{sec:workloads}
Each measurement run drives the GPU with a causal-language-modelling \emph{training} step (forward pass, backward pass, and AdamW optimizer step), so that observed power reflects the full compute graph a training simulator would target.
Input is a single fixed English prompt replicated to the configured batch size, tokenized with the model's own tokenizer, and padded or truncated to the requested context length.
This keeps the data pipeline out of the measurement and isolates on-device compute.

We sweep the following axes:
\begin{itemize}
    \item \textbf{6 GPUs:} NVIDIA A100, L40, L4, Quadro RTX 5000, RTX 4070 Ti, and AMD MI210.
    \item \textbf{3 model families} (Qwen2.5, GPT-2, DialoGPT), with sizes matched to GPU tier:
          commodity (RTX 4070 Ti, Quadro RTX 5000): 0.5B\,/\,124M\,/\,medium;
          mid (L40, L4): 1.5B\,/\,GPT-2-XL\,/\,large\footnote{Qwen2.5-1.5B in float32 with AdamW does not fit in the L4's 24\,GB and is excluded from that cell; the L4's float16 and bfloat16 Qwen runs and the L40's float32 Qwen runs are unaffected.};
          large (A100, MI210): 3B\,/\,GPT-2-XL\,/\,large.
    \item \textbf{3 numerical precisions:} float32, float16, bfloat16\footnote{The Quadro RTX 5000 lacks hardware bfloat16 support.}.
    \item \textbf{7 batch sizes:} 1, 4, 8, 16, 32, 64, 128.
    \item \textbf{2 context windows:} 512 and 2048 tokens.
\end{itemize}
The cross product yields 126 workload configurations per GPU (84 on the Quadro RTX 5000).
Each configuration is repeated four times.

We use the same optimizer (AdamW, learning rate $10^{-4}$, weight decay $0.01$), attention implementation (eager, with flash and memory-efficient SDPA kernels disabled), sampling interval (1\,s of GPU-event time), warmup (5 training steps), cooldown (5\,s between configurations), and single-GPU execution across all runs.
Gradient scaling is disabled across all precisions for consistency.
Clocks, persistence mode, and power limits are left at driver defaults, so devices are free to adjust frequency and voltage as they would in production.
% The resulting variability is bounded: temperature explains under 1.5\,\% of the within-cell residual variance, and the median drift of per-configuration mean power across repeats stays below 0.12\,\%, well under the measurement-noise floor of Table~\ref{tab:noise_floor}.

\subsection{Execution and Telemetry}\label{sec:execution}

Per-step duration is measured with paired CUDA events, and a telemetry sample is emitted roughly once per second of accumulated step time, with in-window GPU~Utilization weighted by per-step duration.

Per-step training FLOPs are approximated as $3\times$ the forward FLOPs reported by CalFLOPS~\cite{calflops}, following the standard 1:2 forward-to-backward ratio for transformer training~\cite{kaplan2020scaling}.
Peak per-device, per-precision FLOPS values were pulled from the official vendor datasheets (Table~\ref{tab:peak_flops}).
The FP32 peaks reflect the Tensor/Matrix-Core path that PyTorch takes under \texttt{set\_float32\_matmul\_precision("high")} with \texttt{allow\_tf32=True} (TF32 Tensor Core on Ampere, FP16 Tensor Core on Turing, BF16 Matrix Core on CDNA2), not IEEE-strict FP32 vector throughput.

\begin{table}[t]
    \centering
    \caption{GPU memory and peak FLOPS (used for MFU normalization) taken from the official vendor datasheets for each device.}
    \label{tab:peak_flops}
    \begin{tabular}{lcccc}
        \toprule
        & \textbf{Memory} & \multicolumn{3}{c}{\textbf{Peak FLOPS (TFLOPS)}} \\
        \cmidrule(lr){3-5}
        \textbf{GPU} & \textbf{(GB)} & \textbf{FP32} & \textbf{FP16} & \textbf{BF16} \\
        \midrule
        NVIDIA A100        & 80 & 156   & 312    & 312    \\
        NVIDIA L40         & 48 & 90.5  & 181.05 & 181.05 \\
        NVIDIA L4          & 24 & 30.3  & 121    & 121    \\
        NVIDIA Quadro RTX 5000 & 16 & 89.2  & 89.2   & ---    \\
        NVIDIA RTX 4070 Ti & 12 & 40.1  & 80.2   & 80.2   \\
        AMD Instinct MI210 & 64 & 181   & 181    & 181    \\
        \bottomrule
    \end{tabular}
\end{table}

For each configuration we sample until the 95\,\% confidence-interval half-width of both MFU and GPU~Utilization falls within 5\,\% of their running mean (i.e.\ $1.96 \cdot \text{SEM}/\bar{x} \le 0.05$), with a minimum of 50 samples and a hard cap of 500.
Power is sampled jointly but excluded from the stopping rule.
It is the smoothest of the three signals, and once both predictors are stable it is stable by extension.

Models are executed with PyTorch~\cite{paszke2019pytorch} and Hugging Face Transformers~\cite{wolf2019huggingface0s}.
Configuration sweeps are managed with Hydra~\cite{Yadan2019Hydra}.
On-device telemetry uses NVML\footnote{\url{https://developer.nvidia.com/management-library-nvml}} on NVIDIA and ROCm\footnote{\url{https://www.amd.com/en/products/software/rocm.html}}~\cite{weakley-2025} on AMD.
FLOP profiling uses CalFLOPS~\cite{calflops}.

\subsection{Aggregation and Metrics}\label{sec:analysis}
Per-sample logs are aggregated to per-configuration means across the four repetitions.
All linear fits in Section~\ref{results} are ordinary least squares of GPU-reported power on the empirical (CalFLOPS-based) MFU, fitted on the per-configuration means.
Only the per-sample error distribution in Figure~\ref{fig:prediction_error} is computed on the raw samples.
$R^2$ and MAPE = $\text{mean}(|y-\hat{y}|/|y|)$ are reported with 95\,\% confidence intervals from a row-level bootstrap with 2000 resamples.

\section{Results}\label{results}
We evaluate how well MFU and GPU Utilization predict GPU power consumption across hardware platforms and workload configurations.
We report predictor--power relationships, error characteristics, and sensitivity to experimental variables.

\begin{table}[H]
    \centering
    \caption{Per-GPU linear MFU-based power fits over all configurations. Errors are 95\,\% bootstrap CIs.}
    \label{tab:per_gpu_mfu_power}
    \begin{tabular}{lccc}
        \toprule
        & \multicolumn{2}{c}{\textbf{MFU}} & \textbf{GPU Util} \\
        \cmidrule(lr){2-3}\cmidrule(lr){4-4}
        \textbf{GPU} & \textbf{$R^2$} & \textbf{MAPE} & \textbf{$R^2$} \\
        \midrule
        NVIDIA A100        & $0.74 \pm 0.03$ & $11.0 \pm 0.7$\,\%           & $0.87$ \\
        NVIDIA L40         & $0.80 \pm 0.03$ & $\phantom{0}7.8 \pm 0.5$\,\% & $0.82$ \\
        NVIDIA L4          & $0.37 \pm 0.03$ & $\phantom{0}3.5 \pm 0.5$\,\% & $0.17$ \\
        NVIDIA Quadro 5000 & $0.84 \pm 0.02$ & $12.5 \pm 1.4$\,\%           & $0.85$ \\
        NVIDIA RTX 4070 Ti & $0.68 \pm 0.06$ & $12.5 \pm 1.1$\,\%           & $0.67$ \\
        AMD MI210          & $0.77 \pm 0.05$ & $13.7 \pm 0.9$\,\%           & ---     \\
        \bottomrule
    \end{tabular}
\end{table}
\begin{figure}[H]
    \centering
    \vspace{-4mm}
    \includegraphics[width=\columnwidth]{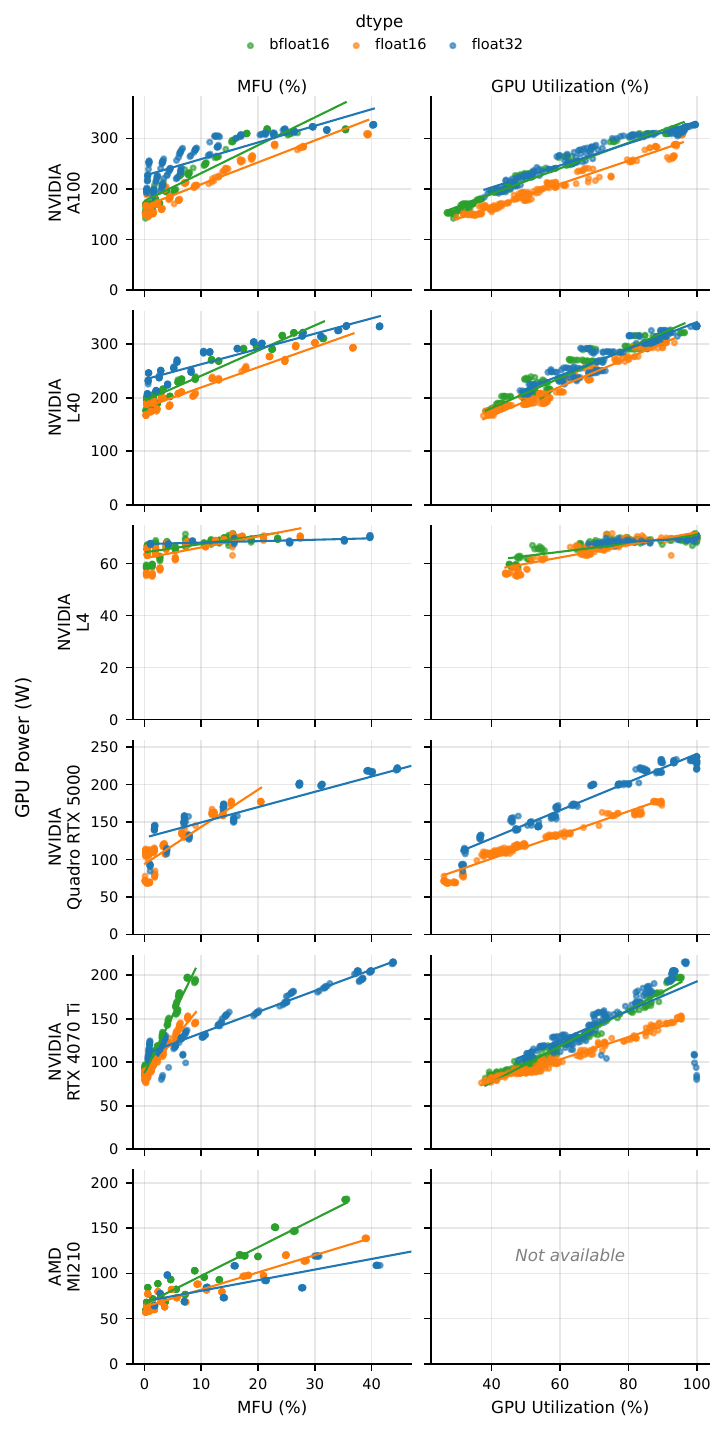}
    \vspace{-6mm}
    \caption{MFU and GPU~Utilization versus GPU-reported power. The MI210 appears only in the MFU panel because its GPU~Utilization signal is effectively binary.}
    \label{fig:Predictors across DTypes}
\end{figure}

\subsection{A Linear MFU-Based Power Model Fits Every GPU}\label{Predictor Performance}
Across all six devices, a linear MFU-based power model fits the GPU-reported power signal.
Table~\ref{tab:per_gpu_mfu_power} reports the per-GPU fits alongside the GPU~Utilization counterpart for context.
The per-GPU MFU fits explain 68--84\,\% of variance in GPU-reported power on training-class NVIDIA cards, dropping to 37\,\% on the inference-class L4, and 77\,\% on the AMD MI210.
Mean absolute percentage errors (MAPE) range from 3.5\,\% on the L4 to 13.7\,\% on MI210.
This is sufficient precision for the ``what-if'' energy comparisons targeted by training simulators.
GPU~Utilization fits are tighter than MFU on NVIDIA hardware (0.67--0.87), but MFU is the only one of the two predictors that yields a fittable model on every device: on the MI210, vendor-reported GPU~Utilization saturates at 100\,\% under any sustained training workload (Section~\ref{Relationship Across Variables}), so no linear power--utilization fit is defined there.

\paragraph{Each device needs its own fit}
The fitted slopes range from 1.21\,W per percent of MFU on the MI210 to 4.49 on the A100, a 3.7$\times$ spread that mirrors the spread in peak per-precision FLOPS across these devices.
Because MFU normalizes throughput by peak FLOPS, equal MFU values correspond to different raw arithmetic work on different chips, and power scales with raw arithmetic work.
The portability claim is therefore that the \emph{model class} (linear, MFU-based) transfers across vendors and precision modes, not that a single slope does.
A simulator targeting a new device needs a one-shot per-GPU calibration sweep, after which power follows from MFU alone.
This is precisely the input simulators already produce as an analytical metric~\cite{agrawal-2025,ozcan-2025}.

\paragraph{Power scales linearly with MFU within each precision}
Holding numerical precision and batch size fixed, MFU explains 98\,\% of power variance on the MI210 (median $R^2 = 0.997$ across 21 cells; mean $0.980$, std $0.038$).
The corresponding NVIDIA medians range from $0.89$ (RTX 4070 Ti) to $0.98$ (A100, Quadro 5000), with the exception of the L4 with $0.43$ due to its low power range.
Figure~\ref{fig:Predictors across DTypes} shows the underlying scatter: within each dtype, power scales near-linearly with MFU on every device, with per-dtype slopes that reflect the peak-FLOPS spread noted above.
The per-GPU $R^2$ values in Table~\ref{tab:per_gpu_mfu_power} are lower because they average over multiple regimes, not because the underlying relationship is noisy.

\paragraph{The L4 is a partial exception}
Its 72\,W TDP leaves only $\sim$10\,W above a 64\,W idle floor (87\,\% of TDP, versus 60--80\,\% on the data-center cards).
With this little signal to fit, $R^2$ drops to 0.37, yet the L4's MAPE is the lowest in the fleet (3.5\,\% per-GPU, 0.6\,\% per cell): in absolute terms the linear model still predicts L4 power accurately, simply because there is little to predict.

\takeaway{A linear MFU-based power model fits every device we measured, including the AMD~MI210 where GPU~Utilization is a binary activity flag and admits no linear fit. The linear relationship holds across vendors and precisions, but each device needs its own fit because MFU is normalized by peak FLOPS. Supporting a new accelerator therefore takes only a one-shot per-GPU calibration sweep.}

\subsection{Conditioning on Dtype and Batch Size}\label{Relationship Across Variables}
We established that a linear MFU-based power model fits every GPU.
Two questions remain: which workload variables shift the slope of that fit, and how far does conditioning on them reduce prediction error before the data themselves run out of signal?

\begin{figure}[t]
    \centering
    \includegraphics[width=\columnwidth]{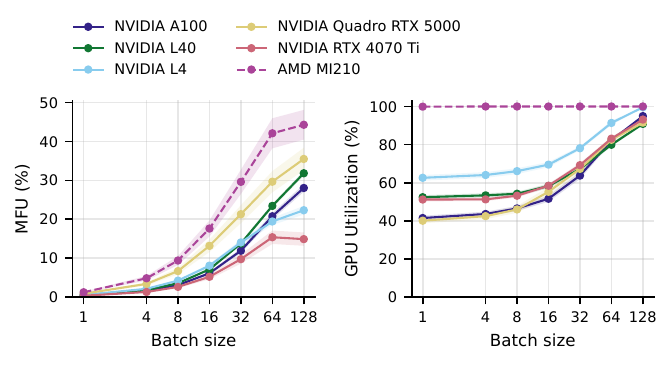}
    \caption{Per-GPU mean MFU (left) and GPU~Utilization (right) as a function of batch size. MFU tracks arithmetic throughput and ramps smoothly with batch on every device. GPU~Utilization rises shallowly on NVIDIA but pins at 100\,\% on the MI210 from batch~1 onward. Shaded bands are $\pm$1\,SEM across configurations.}
    \label{fig:Predictors vs Batch}
\end{figure}

\paragraph{Impact of dtype and batch size}
Of the workload axes we vary, only \emph{numerical precision} and \emph{batch size} materially shift the MFU-power relationship.
Figure~\ref{fig:Predictors across DTypes} shows the per-dtype OLS fits: reduced-precision formats shift the predictor distributions and change the slope of the line, with MFU exhibiting stronger dtype-dependent scaling than GPU~Utilization.
This is expected, since MFU is normalized against a dtype-specific peak FLOPS rate.
Figure~\ref{fig:Predictors vs Batch} illustrates the effect of batch size on MFU and GPU Utilization.
MFU rises smoothly from around 1\,\% at batch~1 to 28--44\,\% at batch~128, tracking arithmetic throughput.
GPU Utilization sits at 40--55\,\% already at batch~1 and saturates near 90\,\% well before MFU does.
On the MI210, GPU Utilization is pinned at 100\,\% across every batch size: ROCm reports utilization from the \texttt{GRBM\_COUNT} counter, which acts as a binary activity flag rather than a throughput proxy and saturates at 100\,\% under any sustained training workload, so the linear fit is undefined on that device.
The tested context window lengths (512 versus 2048 tokens) had no measurable effect on either predictor or on power.

\begin{figure}[t]
    \centering
    \includegraphics[width=\columnwidth]{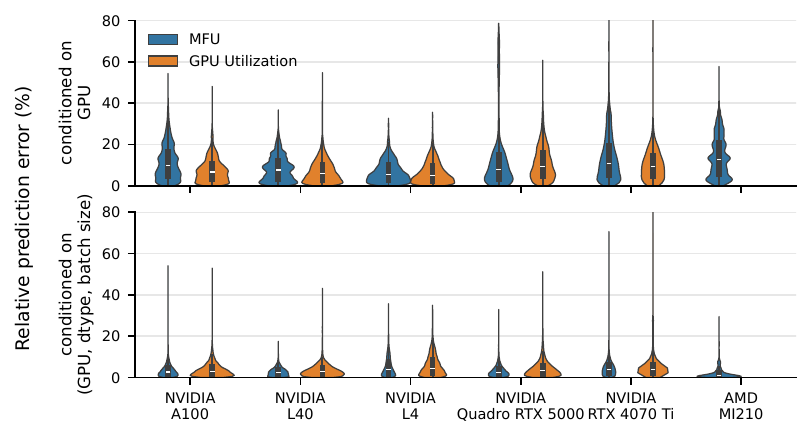}
    \caption{Distribution of per-sample absolute relative error for linear regressions from MFU or GPU Utilization to power draw, across all tested hardware.
             Conditioning on (dtype, batch size) substantially reduces error for both predictors on all devices.}
    \label{fig:prediction_error}
\end{figure}

\begin{figure}[t]
    \centering
    \includegraphics[width=\columnwidth]{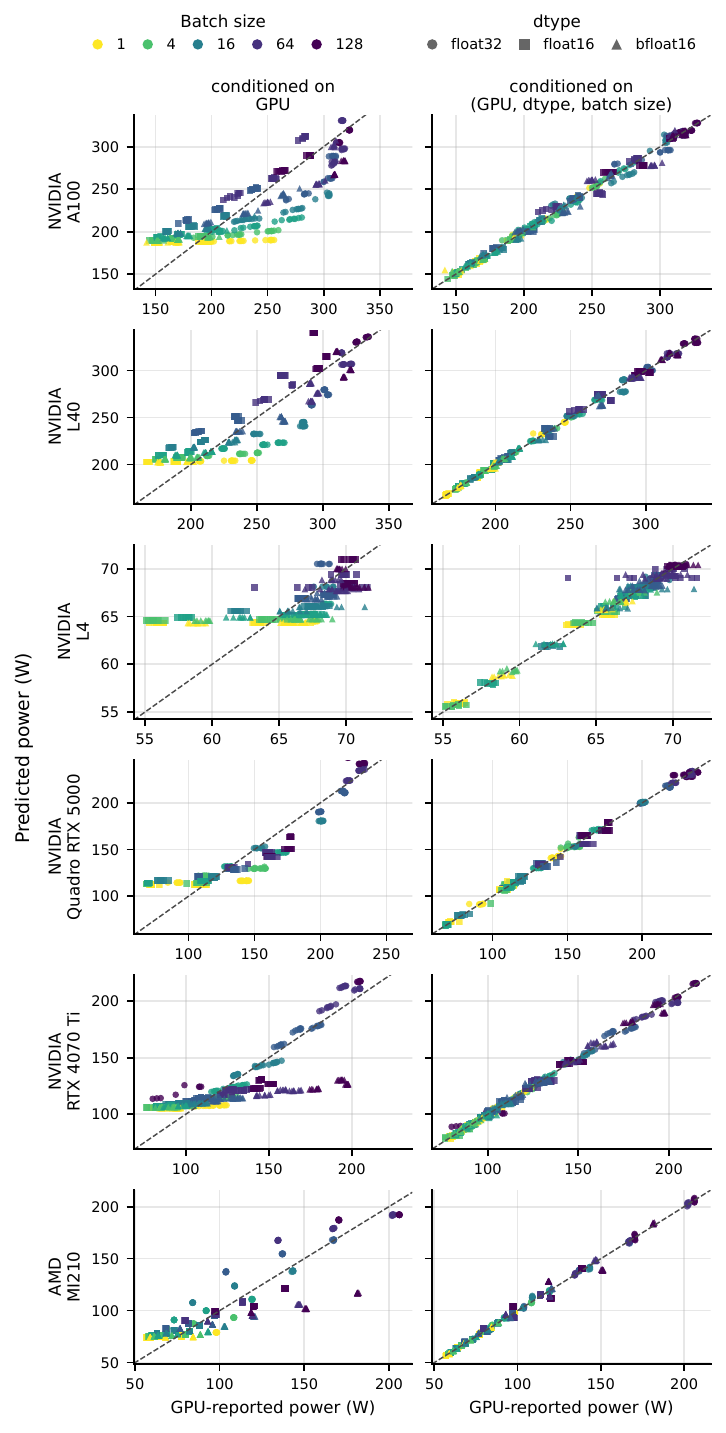}
    \caption{Predicted versus measured GPU power, where the prediction is the fitted value of an OLS regression from MFU to power. Left: a single per-GPU fit; batch-1 configurations (yellow) cluster off the diagonal because the memory-bound regime breaks the single-slope fit. Right: a separate fit per (dtype, batch) cell; predictions collapse onto the diagonal across batch sizes and dtypes.}
    \label{fig:pred_vs_meas_power}
\end{figure}

\paragraph{Conditioning cuts MAPE from $\sim$10\,\% to $\sim$1\,\%}
Fitting a separate linear MFU-power model per (GPU, dtype, batch) cell collapses the prediction error from $\sim$10\,\% per-GPU to $\sim$1\,\% per cell.
Figure~\ref{fig:prediction_error} illustrates the per-sample absolute relative error from MFU or GPU Utilization to power draw.
Figure~\ref{fig:pred_vs_meas_power} compares the two fits directly.
Each point's $y$ value is the power predicted by an OLS regression from MFU, with the left column using a single per-GPU slope and intercept and the right column using a separate slope and intercept fit within each (dtype, batch size) cell of that GPU.
The per-GPU fit (left) shows a clear batch-1 cluster sitting off the diagonal as a flat horizontal band, reflecting the memory-bound regime that the single-slope model cannot capture (Section~\ref{sec:failure-modes}).
Under the (GPU, dtype, batch)-conditioned fits (right) predictions collapse onto the diagonal across all batch sizes and dtypes.
The tail behavior of the per-GPU fit is informative on its own: GPU~Utilization's 95th-percentile error stays below 26\,\% on every NVIDIA device, while MFU can exceed 30\,\% (worst case 57.9\,\% on the Quadro~5000), with the heavier-tail outliers clustering under reduced precision.
The RTX~4070~Ti reverses this ordering: both predictors fit it worst overall, reflecting the workstation environment in which it was measured (Section~\ref{Limitations}).

\paragraph{The conditioned model is at the data limit}
The $\sim$1\,\% MAPE quoted above is an in-sample number.
We investigate whether the remaining error is model misspecification or measurement noise.
Each of the 504 (336 on the Quadro~5000; 392 on the L4) per-GPU configurations is measured across four independent repeats separated by large time gaps.
The cross-repeat standard deviation of mean power therefore bounds any predictor's achievable accuracy.
Table~\ref{tab:noise_floor} compares the per-GPU MFU model's residual SD, the per-(GPU, dtype, batch) model's residual SD, and the cross-repeat measurement-noise SD, each as a percentage of the GPU's mean power.
Conditioning cuts the per-GPU residual by 4--7$\times$ on every device, and on every NVIDIA GPU the conditioned residual is within 1.3--3.7$\times$ of the noise floor.
On the RTX~4070~Ti and L4 it is statistically indistinguishable from it.
The MI210 is the lone outlier in relative terms: its AMD power telemetry is unusually quiet (cross-repeat SD of 0.20\,\% against $\geq$0.35\,\% on every NVIDIA device), but its conditioned residual is in line with the rest of the fleet in absolute terms.

\begin{table}[t]
\centering
\caption{MFU power model residual SD versus the cross-repeat measurement-noise floor, as a percentage of mean GPU power.}
\label{tab:noise_floor}
\small
\begin{tabular}{lccc}
    \toprule
    \textbf{GPU} & \textbf{per-GPU} & \makecell{\textbf{per-(GPU,} \\ \textbf{dtype, batch)}} & \makecell{\textbf{Repeat} \\ \textbf{noise}} \\
    \midrule
        NVIDIA A100 & 12.50\,\% & 2.13\,\% & 1.08\,\% \\
        NVIDIA L40 & \phantom{0}9.02\,\% & 1.29\,\% & 0.35\,\% \\
        NVIDIA L4 & \phantom{0}4.46\,\% & 1.00\,\% & 0.71\,\% \\
        NVIDIA Quadro 5000 & 12.81\,\% & 1.95\,\% & 1.23\,\% \\
        NVIDIA RTX 4070 Ti & 16.46\,\% & 2.21\,\% & 1.66\,\% \\
        AMD MI210 & 17.81\,\% & 2.87\,\% & 0.20\,\% \\
    \bottomrule
\end{tabular}
\end{table}

\takeaway{Within compute-bound regimes, conditioning the MFU power model on dtype and batch size reduces MAPE from $\sim$10\,\% per-GPU to $\sim$1\,\% per (GPU, dtype, batch) cell, and the resulting residual matches the cross-repeat measurement noise on every NVIDIA GPU. In-sample, the remaining error is therefore dominated by sampling noise rather than model misspecification, leaving little headroom for a more elaborate predictor within a calibrated cell.}

\subsection{MFU Underestimates Power in Memory-Bound Regimes}\label{sec:failure-modes}
MFU normalizes throughput by peak FLOPS, so any workload bottlenecked on memory bandwidth rather than arithmetic will execute with low MFU even when power and GPU~Utilization remain high.
Our search grid already brackets this boundary at its low-arithmetic-intensity end: the batch-1 band visible in Figure~\ref{fig:pred_vs_meas_power} is its signature.
Table~\ref{tab:batch1_memory_bound} quantifies the gap.
Going from batch~1 to batch~128, MFU grows by 36--72$\times$ across the six GPUs while power changes by only $1.1$--$2.0\times$.
At batch~1, every device runs below 1.3\,\% MFU yet still draws 49--64\,\% of its batch-128 power---and 90\,\% on the L4, where the idle floor dominates the device's envelope.
A single linear MFU-power slope fit on the broader sweep therefore extrapolates downward through this regime and substantially under-predicts power, with a 70--120\,W absolute error on the data-center cards.

\begin{table}[t]
    \centering
    \caption{Batch-1 (memory-bound) versus batch-128 (compute-bound), averaged over models, precisions, and context windows.}
    \label{tab:batch1_memory_bound}
    \small
    \setlength{\tabcolsep}{4pt}
    \begin{tabular}{lcccc}
        \toprule
        \textbf{GPU} & $\mathbf{MFU_{1}}$ & $\mathbf{MFU_{128}}$ & $\mathbf{P_{1}}$\,(W) & $\mathbf{P_{128}}$\,(W) \\
        \midrule
        NVIDIA A100        & 0.4\,\% & 28.1\,\% & 189 & 307 \\
        NVIDIA L40         & 0.4\,\% & 31.8\,\% & 200 & 315 \\
        NVIDIA L4          & 0.5\,\% & 22.3\,\% & \phantom{0}63 & \phantom{0}70 \\
        NVIDIA Quadro 5000 & 0.8\,\% & 35.5\,\% & 111 & 201 \\
        NVIDIA RTX 4070 Ti & 0.3\,\% & 14.9\,\% & \phantom{0}97 & 169 \\
        AMD MI210          & 1.2\,\% & 44.3\,\% & \phantom{0}72 & 148 \\
        \bottomrule
    \end{tabular}
\end{table}

Batch~1 with 2048-token contexts is the smallest-arithmetic-intensity workload in our sweep, but still well inside the LLM-training scope.
Workloads with even lower arithmetic intensity (decode-style inference, or training with very small models at very long contexts) would extrapolate further outside the regime where a linear MFU-based power model is informative.
A quantitative characterization of that extrapolation, ideally paired with a complementary memory-bandwidth signal, is left to future work.

\takeaway{The MFU-based power model holds inside the compute-bound regime that dominates production LLM training. At its low-arithmetic-intensity edge (batch~1) MFU already collapses below 1.3\,\% while power stays at around 60\,\% of the batch-128 level. A single linear slope cannot serve both regimes, and very-low-batch or decode-style workloads need a complementary memory-bandwidth signal.}

\section{Limitations}\label{Limitations}

In this study, each measurement drives a full forward, backward, and optimizer step (Section~\ref{sec:workloads}), so the fitted MFU-power relation describes training cost.
Autoregressive inference differs along two axes:
First, the decode phase generates one token at a time against a growing KV cache, so per-step arithmetic intensity is low and execution falls into the memory-bound regime where MFU loses predictive power.
Second, production serving additionally introduces request-level dynamics (continuous batching, prefill/decode interleaving, variable prompt and generation lengths) that a fixed training step cannot represent.
We leave an inference evaluation, together with the complementary memory-bandwidth signal it likely requires, to future work.

A second limitation is that all configurations run on a single accelerator, so the fit does not see the inter-device communication component (collectives, pipeline bubbles, FSDP all-gathers) that contributes to power in distributed training.
This is intentional: we isolate the per-device compute-to-power relation, which is the per-node building block that a multi-GPU simulator composes with its own communication model.
Prior work on distributed transformer training already characterizes the scaling and communication side of the problem~\cite{fernandez-2024}.
A cluster simulator reusing our fit must still model communication power and pipeline-bubble idling, and the accuracy of the composed estimate remains to be established.

Lastly, we disabled flash and memory-efficient SDPA kernels in this study and run attention in eager mode across all devices (Section~\ref{sec:execution}).
Production stacks typically rely on fused attention kernels, which raise arithmetic intensity and shift the achieved MFU.
We adopt eager attention so that the MFU-to-power relation is not confounded by kernel-implementation variance across the NVIDIA and AMD toolchains.
We still expect the linear relationship to hold under fused kernels, as any change in slope is exactly what a one-shot per-GPU calibration is supposed to absorb.

\section{Related Work}\label{related work}
GPU power consumption is closely tied to workload-dependent hardware activity.
Previous work has analyzed this relationship through hardware-exposed utilization metrics, telemetry at system scale, and compute-centric abstractions.
However, existing approaches often rely on vendor-specific instrumentation, limiting portability across architectures and preventing direct use in performance modeling or simulation frameworks.

\subsection{Utilization Metrics and Power Correlation}
Elvinger et al.~\cite{elvinger-2025} analyze GPU utilization under kernel colocation and show that coarse metrics such as achieved occupancy fail to capture resource interference, while fine-grained indicators such as instructions per cycle and pipeline utilization are more expressive.
Their study targets performance interference rather than energy, but illustrates that broad vendor-defined counters can obscure the activity that actually matters.

Islam et al.~\cite{islam-2024} bridge utilization and energy more directly, showing through NVML and CUPTI traces that streaming-multiprocessor activity correlates strongly with power draw during compute phases.
Similarly, AccelWattch~\cite{kandiah2021accelwattch} models GPU power from microarchitectural activity counters with high fidelity.
Both depend on architecture-specific instrumentation, however, which limits portability across devices and is unavailable in high-level simulation frameworks.

At system scale, Fernandez et al.~\cite{fernandez-2024} study distributed transformer training and use MFU as a performance indicator, finding that compute efficiency degrades under scaling regimes due to communication overhead.
While MFU is treated as a throughput metric, the work supports the view that compute-centric measures offer a stable abstraction of workload efficiency across configurations.

\subsection{Compute-centric Energy Abstractions}
An alternative approach models energy consumption from computational cost.
Desislavov et al.~\cite{desislavov-2023} estimate inference energy demand by combining model FLOPs with FLOPS-per-Watt efficiency figures across hardware generations.
Their analysis highlights long-term trends in compute growth and hardware efficiency improvements.
This illustrates the appeal of compute-based abstractions as FLOPs are independent from hardware, derivable from model structure, and independent of vendor instrumentation.
However, their estimates rely largely on theoretical efficiency assumptions rather than empirical system-level power measurements, leaving the predictive accuracy of FLOPs-based abstractions uncertain in practical workloads.

\subsection{Portability Challenges in Simulation-Based Modeling}
The limits of hardware-dependent utilization metrics are particularly evident in performance modeling and simulation.
Systems such as Vidur~\cite{agrawal-2025} for inference and existing LLM training simulators and performance-modeling toolkits~\cite{simumax, zhebrak2026llmclustersim, liang-2025} model LLM execution from workload structure and parallelization strategies, without access to hardware-level utilization counters.
Vendor-defined metrics such as GPU Utilization or SM activity therefore cannot feed into the modeling process.
However, training simulators often report MFU as a performance metric, which makes it a natural integration point for the proposed power model.

Özcan et al.~\cite{ozcan-2025} extend Vidur with an MFU-based GPU power model to support energy-aware optimization and for coupling the simulator to co-simulation testbeds like Vessim~\cite{wiesner2024vessim}.
However, their MFU-based power model is not validated against hardware power measurements. We close this gap for compute-bound LLM training.

\section{Conclusion}\label{Conclusion}
In this paper, we evaluated MFU as a portable, software-defined power proxy for LLM training across six NVIDIA and AMD GPUs and nearly 3000 runs.
A linear MFU-power model fits every device in the compute-bound regime, with one fit per (GPU, dtype, batch size) cell.
Because MFU is an analytical quantity derivable from model structure and execution parameters, this result establishes MFU as a predictor for compute-bound LLM training simulation.
Future work should extend the approach to LLM inference simulators such as Vidur~\cite{agrawal-2025}, whose autoregressive decode regime is memory-bound rather than compute-bound.

\appendices
\section{Power Measurement Validation}\label{sec:power-validation}

\begin{figure}[h]
    \centering
    \includegraphics[width=\columnwidth]{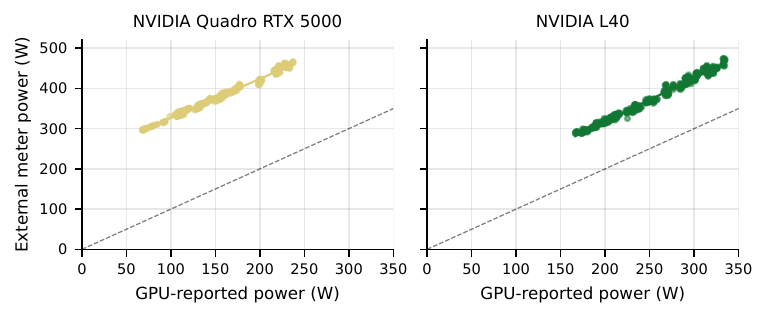}
    \caption{GPU-reported vs. external hardware-based power measurements for NVIDIA L40 and Quadro RTX 5000.}
    \label{fig:Ex. vs Int. Power by Hardware}
\end{figure}

Reliable power measurements are a prerequisite for utilization-based modeling.
For the Quadro RTX 5000 and NVIDIA L40, we recorded external node-level power readings and compared them against GPU-reported power.
As shown in Figure~\ref{fig:Ex. vs Int. Power by Hardware}, GPU-reported power follows a stable, strongly linear relationship with external measurements ($R^2 > 0.99$).
The constant offset represents the host system's base power, which remains invariant under stable GPU load.
This alignment justifies using internal telemetry as the primary modeling target throughout the experiments.

\bibliographystyle{IEEEtran}
\bibliography{sources}

\end{document}